# A Richer Understanding of the Complexity of Election Systems*

Piotr Faliszewski  
Department of Computer Science  
University of Rochester  
Rochester, NY 14627

Edith Hemaspaandra  
Department of Computer Science  
Rochester Institute of Technology  
Rochester, NY 14623

Lane A. Hemaspaandra  
Department of Computer Science  
University of Rochester  
Rochester, NY 14627

Jörg Rothe[†]  
Institut für Informatik  
Heinrich-Heine-Universität Düsseldorf  
40225 Düsseldorf, Germany

September 19, 2006

**Abstract**

We provide an overview of some recent progress on the complexity of election systems. The issues studied include the complexity of the winner, manipulation, bribery, and control problems.

**Key words:** bribery, computational social choice, control, manipulation, voting.

## 1 Introduction

Whether it is "more taste" versus "less filling," "peanut butter" versus "chocolate," or "Bush" versus "Kerry" versus "Nader," people have varying preferences. So it is natural that in life preference aggregation, typically via some voting/election scheme, is a central activity. Within the past few months, the authors of this chapter have seen a department's choice for faculty hiring selected by approval voting and a school's faculty senate election held under single transferable vote, and of course countless actions have been taken under plurality rule and under majority rule. Further, in this modern world of processes and agents, it isn't just people whose preferences must be aggregated. The preferences of computational agents must also be aggregated. Indeed, in both the artificial intelligence and the systems communities a surprisingly broad array of issues have been proposed as appropriate to approach via voting systems. These issues range from spam detection to web search engines to planning in multi-agent systems and much more (see, e.g., [ER91,ER93, PHG00,DKNS01,FKS03]).

---

*Supported in part by DFG grant RO-1202/9-3, NSF grants CCR-0311021 and CCF-0426761, and the Alexander von Humboldt Foundation's TransCoop program. In final form, this survey will appear as a chapter in the book *Fundamental Problems in Computing: Essays in Honor of Professor Daniel J. Rosenkrantz*, to be published by Springer. Author URLs: www.cs.rochester.edu/u/pfali, www.cs.rit.edu/∼eh, www.cs.rochester.edu/u/lane, and ccc.cs.uni-duesseldorf.de/∼rothe.

[†]Work done in part while visiting the University of Rochester.

Thus it is clear that elections are important in both the human and the computer worlds. But why should one study the *complexity* of elections? Although the history of looking at the effect of computational power on decision-making goes quite far back [Sim69], the true genesis of the study of the complexity of elections was a spectacular series of papers by Bartholdi, Orlin, Tovey, and Trick that appeared around 1990 [BTT89b,BTT89a,BO91, BTT92]. One of the insights that naturally drove Bartholdi, Tovey, and Trick [BTT89b] to study complexity issues is that even if an election system has wonderful mathematical properties, if determining who won under the election system is computationally intractable then that system isn't going to be practically useful. Another motivation for studying complexity issues comes from a result known then (the Gibbard–Satterthwaite Theorem), and additional results that have been established since (most notably the Duggan–Schwartz Theorem), showing that every reasonable election system can be manipulated (see [Gib73, Sat75,DS00,Tay05]). So better design of election systems cannot prevent manipulation. Bartholdi, Tovey, and Trick [BTT89a] brilliantly, thrillingly proposed getting around this obstacle by seeking to make manipulation exorbitantly expensive, *computationally*.

The focus areas of those seminal papers from around 1990 were the complexity of the winner problem, the manipulation problem (which regards affecting an election's outcome by changing the votes of voters), and the control problem (which regards affecting an election's outcome by changing the structure of the election—e.g., by adding, deleting, or partitioning voters or candidates). In this chapter, we provide a brief overview of some of the work done on an ongoing project on election complexity that has been pursued over the past decade jointly by the theory groups in Düsseldorf and Rochester. This project has focused on improving the field's understanding of the complexity of winner, manipulation, and control problems, and has also added new directions of inquiry, including the definition and study of election bribery problems. The ultimate goal of the project—which has already been in reasonable part achieved in its manipulation and bribery streams—is to move from simply analyzing individual election systems to finding the source of the complexity of elections. That is, our ultimate goal is to find a simple rule that tells which election systems (perhaps with our focus limited to some broad, important subclass of systems) are computationally simple and which are computationally hard with respect to whichever one of the core questions—winner, manipulation, bribery, or control—is at issue.

By focusing on our own results and interests—though naturally many papers by others are mentioned in the process—we in no way wish to detract from the rest of the enormous body of research being done on related and unrelated topics within the complexity of elections. Indeed, interest in computational social choice theory is at a high level and is still growing, spans fields and countries, and as this is being written the inaugural meeting of a devoted workshop—the (First) International Workshop on Computational Social Choice— is just months away. It is a true, humbling joy to the authors to be part of such a vibrant community with this shared research passion.

Section 2 briefly describes some major election systems. Section 3 studies work showing that the winner problems for Dodgson, Kemeny, and Young elections are complete for parallel access to NP. Section 4 studies work on manipulation and bribery. This work achieves the "simple classification rule" goal mentioned above, and does so on the most important class of election systems—scoring protocols. Section 5 is about electoral control, and studies both the original approach to control and work that extended the control paradigm to the "destructive" case—asking not whether one can make a preferred candidate win, but rather asking whether one can block a despised candidate from winning.

## 2   Elections and Election Systems: Preliminaries

Throughout this chapter, an election, $(C, V)$, will consist of a finite, though arbitrary in size, candidate set $C$ and a finite, though arbitrary in size, voter set $V$. It is legal, though a bit bizarre, for an election to have no candidates or no voters.

Our voters will, unless otherwise specified, be input as a list. Each voter will not be associated with a name, but rather will be input simply via his or her preferences over the candidates. The nature of those preferences depends on the election system. For almost all the election systems discussed in this chapter, each voter is specified as a *tie-free linear ordering of the candidates*, e.g., Bush > Kerry > Nader. We will typically refer to that as the voter's *preference list*. For some systems discussed in this chapter—approval voting and $k$-approval voting—voters are instead specified by *approval vectors*, namely, a vector that for each candidate specifies 1 for approval or 0 for disapproval.

As just mentioned, we generally assume that voters are input as a list. So $V$ might typically be entered as (a natural coding of), for example, the list

$$(\text{Bush} > \text{Kerry} > \text{Nader},\ \text{Nader} > \text{Kerry} > \text{Bush},\ \text{Bush} > \text{Kerry} > \text{Nader}).$$

Note in particular that we do not (except when speaking of *succinct* versions of problems—versions where one can list a preference's multiplicity as a binary number) allow one to specify multiplicities of a given preference other than by listing the same preference multiple times. This nonsuccinct approach to input has been the most common one ever since the seminal work of Bartholdi, Orlin, Tovey, and Trick, and reflects nicely the fact that in real life ballots are cast one per person.

In some problems we do allow voters to be *weighted*, but that is quite different than the succinctness issue. For example, a weight-3 voter is an indivisible object that is quite different from three weight-1 voters (since the latter can potentially be bribed/not-bribed/deleted/etc. separately from and differently than each other).

An *election system* (or *election rule*) is a mapping that takes as input an election $(C, V)$ and outputs a *winner set* $W$ satisfying $\emptyset \subseteq W \subseteq C$. So, in contrast with a social choice function, which typically maps from elections to preference-lists-altered-to-allow-ties, election systems focus completely on separating the candidates into winners and nonwinners.

Nonetheless, the literature on the complexity of election systems is a bit schizophrenic. Some areas of this literature—such as most of the work on the complexity of winner problems—focus on the issue of whether a particular candidate is (or can be made to be) a winner. Other areas of the literature on the complexity of election systems—such as most of the work on the complexity of control problems—focus on the issue of whether a particular candidate is (or can be made to be) a *unique winner*, i.e., to be a winner and to be the only winner. In the literature on manipulation one finds multiple examples of focus on winners and of focus on unique winners, but since the seminal manipulation complexity paper [BTT89a] focused on winners, we will view that as the "traditional" choice for manipulation.

The abovementioned traditional associations between areas and which of "winner" or "unique winner" to study are largely a matter of taste and often date back to choices made in the seminal papers of Bartholdi, Orlin, Tovey, and Trick. One certainly could choose to diverge from them, and researchers sometimes do. For example, the appendix of [HHR06] reanalyzes the complexity of the winnership problems of Dodgson, Kemeny, and Young elections—whose complexity was previously known for the case of "winner"—for the case of "unique winner," and though it takes some work, shows that in each case the complexity of the unique winner problem is the same as the complexity of the winner problem.

Nonetheless, the traditional choices regarding "winner" versus "unique winner" help unify the literature so that papers within a given research stream—say, the study of electoral control—share the same focus and so can be better compared and contrasted. In this chapter, we respect and follow the traditional choices.

Finally, let us briefly define some of the most important election systems. In *approval voting*, each voter is represented by a 0-1 approval vector. To determine the winner, one component-wise adds the vector from each voter, and all candidates who achieve the largest component-wise sum that appears are winners. For each $k \geq 1$, *k-approval voting* is the same as approval voting, except each voter must have exactly $k$ approvals in his or her vote (and thus we must have $\|C\| \geq k$).

The most important class of election systems is the class of *scoring systems* (or *scoring rules* or *scoring protocols*). A scoring system (for $m$-candidate elections) is defined by a *scoring vector* $\alpha = (\alpha_1, \alpha_2, \ldots, \alpha_m)$ satisfying $\alpha_1 \geq \alpha_2 \geq \cdots \geq \alpha_m$. Each voter is represented by a preference list, and the $i$th most preferred candidate on a given voter's preference list gains $\alpha_i$ points due to that voter. Each candidate's point total is the sum of all the points he or she gets. Whoever gets the highest sum is a winner.

*Plurality-rule elections* are based on the family of scoring systems defined by the scoring vectors (), (1), (1,0), (1,0,0), ..., with the vector appropriate to the number of candidates being the one that is used. *Majority-rule elections* technically are not scoring protocols, but rather are the system using the same scoring vector collection as plurality-rule elections but in which a candidate wins exactly if he or she gets strictly more than $\|V\|/2$ points. Note that approval voting technically isn't a scoring protocol or even a one-scoring-vector-per-election-size family of scoring protocols. However, for each $m \geq k$, $m$-candidate $k$-approval voting is a scoring protocol, based on the vector $(\overbrace{1,\ldots,1}^{k},\overbrace{0,\ldots,0}^{m-k})$. Veto elections are based on the family of scoring systems defined by the scoring vectors (), (0), (1,0), (1,1,0), ....

*Condorcet elections* are the system in which to be a winner one must have the property that for each candidate $d$ other than oneself it must hold that one is preferred to $d$ by strictly more than half the voters (i.e., one wins all head-on-head majority-rule beauty contests).

Such a candidate is called a *Condorcet winner*.

## 3  Complexity of Winning: Dodgson's 1876 Election System

> "I suspect that one of the March Hare, the Hatter, and the Dormouse is guilty," said the Queen, "though I don't know which one is. Thus we—the Duchess, you (Alice), and I—will vote on this matter. Off with the head of any one of the March Hare, the Hatter, and the Dormouse who is preferred to each of the others in pairwise majority-rule contests on whom to execute!"
>
> "You cannot do that," Alice screamed, totally horrified. The Queen replied angrily, "Yes, I *can* do that. This is a rational society where people vote rationally on issues... such as which of those three to behead." And she pointed again to the Hatter, the March Hare, and the Dormouse (who had fallen asleep). "My preference list as to whom to behead is Hatter (I hate him) > March Hare > Dormouse (he is so cute). So shall it be off with the Hatter's head?"
>
> "Not so fast," said the Duchess. "My preference list for whom to behead is March Hare (oh, to rid this world of those creepy long ears!) > Dormouse > Hatter." Suddenly turning to Alice, she asked, "What's your vote?" Alice timidly replied, "If I absolutely must give a list, then my preference list as to whom to behead is Dormouse > Hatter > March Hare."
>
> "Ha!" exclaimed the Queen. "The Hatter is preferred to the March Hare for execution by two to one. Off with the Hatter's head!" "No," replied the Duchess, "the Dormouse is preferred to the Hatter for execution by two to one." "Then off with the Dormouse's head!" cried the Queen. "No," said the Duchess, "the March Hare is preferred to the Dormouse for execution by two to one." "*Then kill the March Hare!*" screamed the Queen, now really quite upset. "Need I remind you," said Alice, "that the Hatter is preferred to the March Hare for execution by two to one? So no one shall be beheaded."
>
> The Queen summarized, "This makes me dizzy. In our rational society, each of the three of us had noncyclic (rational) preferences over these three candidates. And yet when we aggregated our preferences under pairwise majority-rule contests, our societal preference was strictly cyclic: March Hare > Dormouse, Dormouse > Hatter, Hatter > March Hare. Our rational individual preferences aggregated to an irrational societal preference. Since as the Queen I represent the society, perhaps the only fitting penalty is 'Off with *my* head!'"

Lewis Carroll—whose real name was Charles L. Dodgson and who not only was the author of wonderful children's books but also was a mathematician—noticed the same issue the Queen just reached: Rational individual preferences (even with ties not allowed) can aggregate, under pairwise majority-rule contests, to an irrational societal preference. That is, Dodgson rediscovered what is known today as the Condorcet paradox, though he most likely was unaware (see [Bla58, pp. 193–194]) of Condorcet's much earlier work [Con85]. Note that every election instance having this type of strict cycle over all the candidates in the aggregate behavior is a case where there is no Condorcet winner (though not every election instance having no Condorcet winner is a case of this type of strictly cyclic aggregate behavior). In his 1876 essay "A Method of Taking Votes on More than Two Issues," Dodgson [Dod76] proposed an election system that respects Condorcet winners when they exist, and when they don't exist reflects the philosophy that whoever is "closest" to being a Condorcet winner should be declared a winner. In Dodgson's system, given an election $(C, V)$, each candidate $c \in C$ is assigned a score (denoted by $dscore_{(C,V)}(c)$, and we will write just $dscore(c)$ when the election is clear from context): $dscore(c)$ equals the smallest number of sequential exchanges (called "switches" henceforward) of adjacent candidates in the voters' preference lists that suffices to make $c$ a Condorcet winner. Whoever has the lowest Dodgson score wins in Dodgson's system. When a Condorcet winner exists, he or she is clearly the unique candidate with Dodgson score zero and thus is a (indeed, *the*) Dodgson winner as well.

In the above example, there is no Condorcet winner but switching the Hatter and the Dormouse in Alice's preference list yields Hatter > Dormouse > March Hare. So the Hatter now defeats both the March Hare and the Dormouse by two to one in pairwise majority-rule contests and thus is now a Condorcet winner (in the election for the questionable privilege of being beheaded). So in the above example $dscore(\text{Hatter}) = 1$. Similarly, $dscore(\text{March Hare}) = dscore(\text{Dormouse}) = 1$. If there is no Condorcet winner, Dodgson winners are not necessarily unique, though at least one Dodgson winner always exists in Dodgson elections (except when $\|C\| = 0 \vee (\|V\| = 0 \wedge \|C\| \neq 1)$ holds).

How hard is it to determine whether a distinguished candidate is a Dodgson winner of a given election? Bartholdi, Tovey, and Trick [BTT89b] crisply, naturally formalized this problem as follows and also defined two related problems, the scoring and ranking problems for Dodgson elections. A Dodgson triple $(C, c, V)$ consists of an election $(C, V)$ and a distinguished candidate $c \in C$.

**Name:** Dodgson-winner.
**Given:** A Dodgson triple $(C, c, V)$.
**Question:** Is $c$ a Dodgson winner in $(C, V)$, i.e., does $dscore(c) \leq dscore(d)$ hold for each $d \in C$?

**Name:** Dodgson-score.
**Given:** A Dodgson triple $(C, c, V)$ and a nonnegative[1] integer $k$.
**Question:** Is it the case that $dscore(c) \leq k$?

**Name:** Dodgson-ranking.
**Given:** An election $(C, V)$ and two distinguished candidates from $C$, $c$ and $d$.
**Question:** Is it the case that $dscore(c) \leq dscore(d)$?

Bartholdi, Tovey, and Trick [BTT89b] proved that Dodgson-score is NP-complete and that Dodgson-ranking and Dodgson-winner are NP-hard. For the latter two problems Bartholdi, Tovey, and Trick left open whether their lower bounds were optimal, i.e., whether their Dodgson-ranking and Dodgson-winner NP-hardness results could be strengthened to NP-completeness or, alternatively, whether their NP-hardness lower bounds could be raised, ideally to some matching upper bound. These open questions were resolved by the following result of Hemaspaandra, Hemaspaandra, and Rothe [HHR97a].

**Theorem 3.1 ([HHR97a])** Dodgson-ranking *and* Dodgson-winner *are* $\Theta_2^p$-*complete.*

$\Theta_2^p$ here represents, as is standard, a particular level of the polynomial hierarchy. $P_\parallel^{NP}$ is the level of the polynomial hierarchy formed by the class of problems solvable by parallel (i.e., truth-table) access to NP. Quite early, Papadimitriou and Zachos [PZ83] studied $P^{NP[\log]}$, the class of problems solvable by $\mathcal{O}(\log n)$ sequential (i.e., Turing) queries to NP. However, it is now known that $P^{NP[\log]}$ and $P_\parallel^{NP}$ are equal [Hem89], and the class they each define is often referred to as the $\Theta_2^p$ level of the polynomial hierarchy. There are surprisingly many characterizations of $\Theta_2^p$ (see [Wag90])—a tribute to its robustness under definitional variation. From the definitions, $\Theta_2^p$ is easily seen to be related to other polynomial hierarchy levels as follows: $\text{NP} \cup \text{coNP} \subseteq \Theta_2^p \subseteq P^{NP} \subseteq NP^{NP} \cap coNP^{NP}$.

The remainder of this section is mainly devoted to sketching the proof of Theorem 3.1. Our proof sketch proceeds via a series of lemmas. The general proof structure is shown in Figure 1.

The $\Theta_2^p$ upper bounds for Dodgson-ranking and Dodgson-winner are easy to see. Dodgson-ranking, for example, is in $\Theta_2^p$ via the simple algorithm that, given an instance $((C, V), c, d)$, uses the NP oracle Dodgson-score to compute $dscore(c)$ and $dscore(d)$ by asking in parallel all plausible values of those scores, and by doing so will discover whether $dscore(c) \leq dscore(d)$.

---

[1] Both [BTT89b] and [HHR97a] have "positive" here rather than "nonnegative," but it is easy to see that the NP-completeness of this problem is unaffected by that word change (basically because the $k = 0$ case can be tested for in polynomial time).

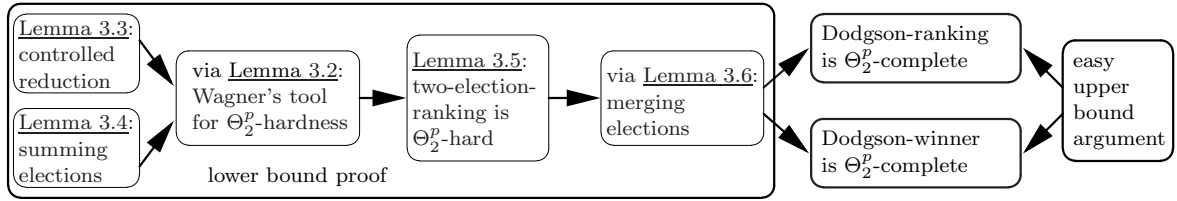

Figure 1: Proof structure for Theorem 3.1.

Regarding the lower bounds, we show Dodgson-ranking and Dodgson-winner $\Theta_2^p$-*hard* by proving many crucial properties of/operations on Dodgson elections to be *easy*, i.e., we provide polynomial-time algorithms for them. Although this may at first seem counterintuitive, it in fact is natural: Showing that a problem $A$ is $\leq_m^p$-hard for a class reflects not just that $A$ has the power to solve all the sets from the class, but also that $A$ is so nicely and simply structured that a polynomial-time many-one reduction from each set in the class can tap into that power. So $\leq_m^p$-hardness is itself about simplicity and the power of polynomial-time transformations.

In our setting, some of the easiness results we obtain will be used hand-in-hand with a $\Theta_2^p$-hardness tool of Wagner ([Wag87], see also the surveys [HHR97b, RR06]), stated as Lemma 3.2 below. Along the lines of the previous paragraph, to link our problems to this tool we must explore the properties of Dodgson elections and in particular how to resculpt them via efficient algorithms.

**Lemma 3.2 ([Wag87])** *Let $A$ be an NP-complete set, and let $B$ be any set. Then $B$ is $\Theta_2^p$-hard if there is a polynomial-time function $f$ such that, for all $k \geq 1$ and all $x_1, \ldots, x_{2k} \in \Sigma^*$ satisfying $\chi_A(x_1) \geq \chi_A(x_2) \geq \cdots \geq \chi_A(x_{2k})$, it holds that $\|\{i \mid x_i \in A\}\| \equiv 1 \pmod{2} \iff f(x_1, \ldots, x_{2k}) \in B$.*

To exploit Lemma 3.2, we have to do much groundwork. Bartholdi, Tovey, and Trick [BTT89b] proved Dodgson-score NP-hard via a reduction from exact-cover-by-three-sets. However, their reduction does not have the properties needed to exploit Lemma 3.2. In contrast, one can achieve these properties by constructing a reduction to Dodgson-score that starts from the well-known NP-complete problem three-dimensional-matching (see Garey and Johnson [GJ79] for specifics on three-dimensional-matching), which we will for brevity henceforward refer to as 3DM. This reduction has the property—which is vastly more restrictive than what is needed merely to achieve a vanilla many-one reduction in this case—that when it reduces to a question about whether a certain candidate has score at most $k$ in a given election, it will always be the case that that candidate's true score in that election is either $k$ or $k + 1$.

**Lemma 3.3 ([HHR97a])** *There is a polynomial-time function $f$ that reduces* 3DM *to* Dodgson-score *in such a way that, for each $x \in \Sigma^*$, $f(x) = ((C, c, V), k)$ is an instance of* Dodgson-score *with an odd number of voters and this instance has the property that: (a) if $x \in$* 3DM *then $dscore(c) = k$ and (b) if $x \notin$* 3DM *then $dscore(c) = k + 1$.*

Next, Lemma 3.4 shows how to "sum" Dodgson triples in such a way that the Dodgson score of the "sum" equals the sum of the Dodgson scores of the given Dodgson triples.

**Lemma 3.4 ([HHR97a])** *There is a polynomial-time function dodgsonsum such that, for all $\ell$ and for all Dodgson triples $(C_1, c_1, V_1), (C_2, c_2, V_2), \ldots, (C_\ell, c_\ell, V_\ell)$ each having an odd number of voters, $dodgsonsum((C_1, c_1, V_1), (C_2, c_2, V_2), \ldots, (C_\ell, c_\ell, V_\ell)) = (C, c, V)$ is a Dodgson triple with an odd number of voters and satisfies $dscore_{(C,V)}(c) = \sum_{1 \le j \le \ell} dscore_{(C_j, V_j)}(c_j)$.*

We now define an ancillary problem that is closely related to Dodgson-ranking and Dodgson-winner.

**Name:** two-election-ranking (2ER, for short).
**Given:** Two Dodgson triples, $(C, c, V)$ and $(D, d, W)$, with $c \ne d$ and $\|V\|$ odd and $\|W\|$ odd.
**Question:** Is it the case that $dscore_{(C,V)}(c) \le dscore_{(D,W)}(d)$?

Lemmas 3.2, 3.3, and 3.4 can be used (together with about two pages of additional argumentation) to obtain Lemma 3.5. Note that 2ER plays the role of the set $B$ in Lemma 3.2, and 3DM plays the role of that lemma's NP-complete set $A$. Note also that 2ER is in $\Theta_2^p$, so Lemma 3.5 implies that 2ER is $\Theta_2^p$-complete.

**Lemma 3.5 ([HHR97a])** 2ER *is $\Theta_2^p$-hard.*

Finally, Lemma 3.6 shows how to merge two Dodgson elections into a single Dodgson election in a very careful way such that a number of useful properties are achieved. Using this lemma we can transfer 2ER's $\Theta_2^p$-hardness to both Dodgson-ranking and Dodgson-winner.

One can think of Lemma 3.6, informally, as akin to a "double-exposure" photograph: Our merged election retains and reflects important information about both its underlying elections.

**Lemma 3.6 ([HHR97a])** *There are polynomial-time functions merge and merge′ such that, for all Dodgson triples $(C, c, V)$ and $(D, d, W)$ for which $c \neq d$ and both $\|V\|$ and $\|W\|$ are odd, there exist $\widehat{C}$ and $\widehat{V}$ such that*

1. $merge((C, c, V), (D, d, W)) = ((\widehat{C}, \widehat{V}), c, d)$ *is an instance of* Dodgson-ranking,

2. $merge'((C, c, V), (D, d, W)) = (\widehat{C}, c, \widehat{V})$ *is an instance of* Dodgson-winner,

3. $dscore_{(\widehat{C},\widehat{V})}(c) = dscore_{(C,V)}(c) + 1$,

4. $dscore_{(\widehat{C},\widehat{V})}(d) = dscore_{(D,W)}(d) + 1$, *and*

5. *for each* $e \in \widehat{C} - \{c, d\}$, $dscore_{(\widehat{C},\widehat{V})}(c) < dscore_{(\widehat{C},\widehat{V})}(e)$.

Space is too tight to cover in detail here the nine pages of proofs for Lemmas 3.3 through 3.6 and Theorem 3.1. But to give the reader at least some flavor of how the proofs work, we illustrate by an example the construction used for proving Lemma 3.6. Let the Dodgson triples $(C, c, V)$ and $(D, d, W)$ be given, where $C = \{a, b, c\}$ and $D = \{d, e, f\}$, $V$ contains three preference lists, $c > b > a$, $a > c > b$, and $b > a > c$, and $W$ contains one preference list, $f > e > d$. Clearly, $dscore_{(C,V)}(c) = 1$ and $dscore_{(D,W)}(d) = 2$. Now construct the election $(\widehat{C}, \widehat{V})$, which is part of the output of the functions *merge* and *merge′*, as follows. The candidate set is $\widehat{C} = C \cup D \cup S \cup T$, where $S$ and $T$ are sets of so-called separating candidates.[2] Voter set $\widehat{V}$ consists of the following preference lists:

1. $c > b > a > \overrightarrow{S} > e > f > \overrightarrow{T} > d$,

2. $a > c > b > \overrightarrow{S} > e > f > \overrightarrow{T} > d$,

3. $b > a > c > \overrightarrow{S} > e > f > \overrightarrow{T} > d$,

4. $f > e > d > a > b > \overrightarrow{T} > c > \overrightarrow{S}$,

5. $d > e > f > a > b > \overrightarrow{T} > c > \overrightarrow{S}$,

---

[2] To make the proof of Lemma 3.6 work in general, $S$ and $T$ have to be chosen sufficiently large. In this toy example, however, all properties hold even if $\|S\| = \|T\| = 0$, so all separating candidates could be dropped here.

6. $d > c > \overleftarrow{T} > e > f > a > b > \overrightarrow{S}$,

7. $d > c > \overleftarrow{T} > e > f > a > b > \overrightarrow{S}$, and

8. $c > d > \overrightarrow{T} > e > f > a > b > \overrightarrow{S}$,

where $\overrightarrow{S}$ (respectively, $\overrightarrow{T}$) represents the candidates of $S$ (respectively, $T$) in some fixed order, and (to avoid interference regarding property 5 of the lemma) $\overleftarrow{T}$ represents the candidates of $T$ in the order that reverses their order in $\overrightarrow{T}$. The first three voters in $\widehat{V}$ simulate $V$, the fourth voter simulates $W$, and the remaining voters are so-called normalizing voters. Properties 1 and 2 of Lemma 3.6 are immediate.

For properties 3 and 4 of Lemma 3.6, let us determine the Dodgson scores of candidates $c$ and $d$ in $(\widehat{C}, \widehat{V})$. Note that one switch in, say, the second voter of $\widehat{V}$ (which is one of the voters simulating $V$) gives the new preference list $c > a > b > \overrightarrow{S} > e > f > \overrightarrow{T} > d$, and one switch in, say, the sixth voter of $\widehat{V}$ (a normalizing voter) gives $c > d > \overleftarrow{T} > e > f > a > b > \overrightarrow{S}$. By these two switches, $c$ has become a Condorcet winner, so $dscore_{(\widehat{C},\widehat{V})}(c) \leq 2$. But since $c$ needs to gain one vote in $(\widehat{C}, \widehat{V})$ against each of $a$ and $d$ to defeat all candidates by a strict majority, no single switch in the preference lists of $\widehat{V}$ can make $c$ a Condorcet winner, so $dscore_{(\widehat{C},\widehat{V})}(c) \geq 2$. Thus $dscore_{(\widehat{C},\widehat{V})}(c) = 2 = dscore_{(C,V)}(c) + 1$. Similarly, two switches in the fourth voter of $\widehat{V}$ (which simulates $W$) gives the new preference list $d > f > e > a > b > \overrightarrow{T} > c > \overrightarrow{S}$, and one switch in the eighth voter of $\widehat{V}$ (a normalizing voter) yields $d > c > \overrightarrow{T} > e > f > a > b > \overrightarrow{S}$. By these three switches, $d$ has become a Condorcet winner, so $dscore_{(\widehat{C},\widehat{V})}(d) \leq 3$. Again, since $d$ needs to gain one vote in $(\widehat{C}, \widehat{V})$ against each of $c$, $e$, and $f$ to defeat all candidates by a strict majority, no two switches in the preference lists of $\widehat{V}$ can make $d$ a Condorcet winner, so $dscore_{(\widehat{C},\widehat{V})}(d) \geq 3$. Thus $dscore_{(\widehat{C},\widehat{V})}(d) = 3 = dscore_{(D,W)}(d) + 1$. Property 5 uses similar arguments.

As mentioned earlier, Dodgson's system respects the Condorcet winner when a Condorcet winner exists. Since the notion of Condorcet winner is widely considered central and important, election systems with this property have been intensely studied (see Fishburn [Fis77]). Some other examples of election systems respecting the notion of Condorcet winner are those of Young [You77] and Kemeny [Kem59,KS60]. In Young's system, whoever can be made a Condorcet winner by removing the smallest number of voters wins. Rothe, Spakowski, and Vogel [RSV03] proved that the winner problem for Young elections is $\Theta_2^p$-complete, via a reduction from the problem maximum-set-packing-compare. Kemeny's winners are defined via the notion of a "Kemeny consensus." Each ranking of the candidates (with ties allowed) that is "closest" to the given preference lists of the voters with respect to a certain distance function is a Kemeny consensus. A candidate is a winner in a Kemeny election if there exists some Kemeny consensus in which that candidate is a winner (a highest-ranked candidate, though possibly tied for that position). Hemaspaandra, Spakowski, and Vogel [HSV05] proved that the winner problem for Kemeny elections is $\Theta_2^p$-complete, via a reduction from the problem feedback-arc-set-member.

The three $\Theta_2^p$-completeness results discussed above pinpoint the complexity of the winner problems for Dodgson, Young, and Kemeny elections. Winners in these three systems are not necessarily unique and these three winner problems ask whether a given candidate is *a winner*. However, as mentioned earlier, Hemaspaandra, Hemaspaandra, and Rothe [HHR06] have shown that the unique winner problems for Dodgson, Young, and Kemeny elections are $\Theta_2^p$-complete as well.

$\Theta_2^p$-completeness suggests that the relevant problem is far from being efficiently solvable, and there are many ways in which completeness for this higher level of the polynomial hierarchy speaks more powerfully than would completeness for its kid brother, NP [HHR97b]. Since checking whether a given candidate has won should be in polynomial time in any system to be put into actual use, these results show that Dodgson, Young, and Kemeny elections are unlikely to be useful in practice. However, we note that winnership in "homogeneous" Young elections can indeed be tested in polynomial time via integer linear programming [You77], that Dwork et al. [DKNS01] have proposed an efficient heuristic (called "local Kemenization") regarding the Kemeny winner problem, and that Dodgson winners can be determined efficiently (a) for elections with a bounded number of candidates or voters [BTT89b], (b) in Fishburn's [Fis77] homogeneous variant of Dodgson elections [RSV03], and (c) with a guaranteed high frequency of success under a simple greedy heuristic [HH06].

# 4 Complexity of Manipulation and Bribery: Scoring Systems and Dichotomy Theorems

The previous section studied the complexity of winner problems for certain election systems. We in this section turn to electoral problems that formalize attempts to influence an election's outcome for the case of a group of manipulative voters (who strategically change their preference lists) and for the case of having someone trying to bribe voters to change their preference lists. The next section studies attempts to influence elections via altering their structure. For a given election system, one would naturally most hope to find that it has an easy winner problem but that it resists electoral manipulation, bribery, and control.

Unfortunately, voters may often be tempted to cast their votes not according to their true preferences but rather insincerely, based on strategic considerations. Consider the following example. Our voting system is the Borda count, a family of scoring protocols that for $m$ candidates uses the scoring vector $\alpha = (m-1, m-2, \ldots, 0)$. We have three candidates, $a$, $b$, and $c$, and eleven voters, where five voters have preference list $a > b > c$, five voters have preference list $b > a > c$, and one voter has preference list $c > a > b$. Under the Borda count scoring procedure, candidate $a$ receives 16 points, candidate $b$ receives 15 points, and candidate $c$ receives 2 points. So $a$ is the unique winner. However, from the point of view of voters with (true) preference list $b > a > c$, it might be tempting to report $b > c > a$ instead. This way they might actually make $b$ win. Namely, if all voters whose sincere preference list is $b > a > c$ were to instead cast $b > c > a$ as their votes (and all other votes were to remain unchanged), then $b$ would become the unique winner.

Of course, we would like to have election systems that cannot be manipulated. Unfortunately, a powerful line of work shows that all practically useful systems operating on three or more candidates are open to manipulation. In particular, the Gibbard–Satterthwaite Theorem [Gib73, Sat75] shows that for each nondictatorial election system that always selects exactly one winner and in which the candidate set is of size at least three and in which for each candidate there is some set of votes that make that candidate a winner, there exists some situation in which a single strategic voter has an incentive to vote insincerely. The Duggan–Schwartz Theorem ([DS00], see also [Tay05]) obtains an analog of this for the model in which—as in this paper—the winner set is some subset (possibly empty, possibly nonstrict) of the candidate set.

Although manipulation cannot be absolutely precluded in any reasonable election system on three or more candidates, Bartholdi, Orlin, Tovey, and Trick [BTT89a, BO91] ingeniously proposed to at least make it computationally prohibitive—e.g., NP-hard—for a manipulator (or in later work by others, a coalition of manipulators) to figure out whether (and how) his/her/their vote(s) can be modified so as to make a given candidate win. They found systems *vulnerable to manipulation* (i.e., one can tell in polynomial time whether and how a given candidate can be turned into a winner) and they found systems *resistant to manipulation* (i.e., systems for which the manipulation problem is NP-hard). This line of research has been very actively pursued ever since (see, as just a few of the many examples on or related to this, [CS02, CLS03, EL05, HH05, FHH06, PRZ06, PR06]). As mentioned in the introduction, complexity issues for electoral problems (in particular, manipulation problems) are particularly important in these modern times when preference aggregation is used in multi-agent systems, distributed computing, and internet applications.

A problem closely related to manipulation is bribery. In (constructive) manipulation, a group of manipulators wants to, by setting their own preference lists, have a given candidate end up a winner. (Destructive manipulation, which analogously seeks to have a given candidate end up a nonwinner, has also been studied [CS02,CLS03]. This section deals with the constructive case only.) Bribery is related to manipulation, except that now we look at elections from the point of view of an outside agent who wants to make some candidate win and who has some budget to bribe voters to change their votes. We now formally define these problems for a given election system $\mathcal{E}$.

**Name:** $\mathcal{E}$-manipulation.
**Given:** A set $C$ of candidates, a set $V$ of nonmanipulative voters, a set $S$ of manipulative voters with $V \cap S = \emptyset$, and a distinguished candidate $c \in C$.
**Question:** Is there a way to set the preference lists of the voters in $S$ such that, under election system $\mathcal{E}$, $c$ is a winner of election $(C, V \cup S)$?

Instead of the plain $\mathcal{E}$-manipulation problem presented above, we often are interested in the $\mathcal{E}$-weighted-manipulation problem in which voters, both in $V$ and $S$, have weights. (In this version of the problem, the set $S$ is usually represented simply as a list of weights of the manipulators.)

The bribery problem for election system $\mathcal{E}$ is defined similarly, except that in bribery the set of voters who can change their preference lists is not part of the input. Intuition might say that bribery thus is more difficult than manipulation, but in fact that is not necessarily the case. We will see (and [FHH06] has a full treatment) that some bribery problems are NP-complete yet their manipulation analogs are in P, some bribery problems are in P yet their manipulation analogs are NP-complete, and sometimes both problems are equally complex.

**Name:** $\mathcal{E}$-bribery.
**Given:** A set $C$ of candidates, a set $V$ of voters, a distinguished candidate $c \in C$, and a nonnegative integer $k$.
**Question:** Is it possible to change the preference lists of at most $k$ voters such that, under election system $\mathcal{E}$, $c$ is a winner of election $(C, V)$?

Bribery also has its weighted version, $\mathcal{E}$-weighted-bribery. In addition, bribery can come in a few other natural flavors. In particular, in $\mathcal{E}$-$bribery we associate each voter with a

price tag and interpret the integer $k$ as a budget, and we ask whether it is possible to make $c$ a winner by changing the preference lists of voters whose total price does not exceed $k$. If the voters have both price tags and weights then we call the problem $\mathcal{E}$-weighted-\$bribery. Sometimes we also put special restrictions on how to represent weights or prices, and will indicate such restrictions by subscripts.

Throughout this section we will be interested only in manipulation and bribery problems that ask about making the distinguished candidate a winner (as opposed to a unique winner—though we mention in passing that, very often, analogous results hold for both cases, see [FHH06]).

In the remainder of this section we present the flavor of some recently obtained results on manipulation and bribery. We point out that in this section we discuss manipulation and bribery together as if they had been developed at the same time, but this is not really the case. The complexity of manipulation has long been studied, but the study of the complexity of bribery in elections started very recently. Nonetheless, the relationships between these two families of problems are very interesting and natural, and we feel that it is more instructive to present these results together.

One approach to the study of manipulation and bribery would be to study these questions one election system at a time. Let us for a moment do that—by focusing on one of the best-known and most popular election systems, plurality-rule elections— though we will soon seek to wrap many of these results about plurality-rule elections into a broader framework that allows insights into hardness to span many systems. Recall from Section 2 that a plurality-rule election with $m$ candidates is described by the scoring vector $(1, \overbrace{0, \ldots, 0}^{m-1})$, so only the top candidate in each preference list matters.[3] First note that, not surprisingly, plurality-rule elections are easy to manipulate.

**Theorem 4.1 ([BTT89a])** plurality-manipulation *and* plurality-weighted-manipulation *are in* P.

---

[3]There is one issue that one should be aware of when discussing families of scoring protocols such as plurality, veto, or Borda count. Formally, each scoring protocol regards only a fixed, constant number of candidates. When we refer to names such as plurality, veto, and Borda count we typically have in mind the whole family of protocols that involves one incarnation of a particular scoring protocol for each candidate set multiplicity. Thus when we discuss the complexity of plurality-rule elections here, we actually give polynomial-time algorithms that are polynomial both in the number of voters and candidates.

In contrast, when we are discussing scoring protocols in general, we as is standard consider a particular scoring vector and thus a fixed number of candidates. This makes NP-completeness results stronger and polynomial-time membership results weaker. However, we note in passing that Hemaspaandra and Hemaspaandra [HH05, Section 3] provide a formalism and a dichotomy (i.e., complete classification) result for manipulation under uniform *families* of scoring protocols.

To prove this theorem, it is enough to observe that if the manipulators want $c$ to become a winner then they should vote for $c$. ([BTT89a] discusses only the unweighted manipulation problem, but clearly weights do not change anything here.)

Now that we know the complexity of manipulating plurality-rule elections, it is natural to ask about the complexity of bribery within such elections. Does the fact that in bribery one has to find some group of voters whose votes are to be changed make the problem more difficult? The answer is no.

**Theorem 4.2 ([FHH06])** plurality-bribery *is in* P.

A greedy algorithm works for plurality-bribery. If we want to make $c$ a winner by bribing at most $k$ voters, we first test whether $c$ is a winner already. If so, we are done. Otherwise, if $k > 0$ then we pick one of the current winners, bribe one of his or her voters to vote for $c$, decrease $k$ by one (if $k$ becomes negative, this means that we used too many bribes and so $c$ cannot be made a winner), and loop back to testing whether $c$ is a winner already.

This algorithm is very simple and natural. Unfortunately, it does not work for the weighted case or for the case of priced voters. In the weighted case it is not always clear whether one should first bribe the heaviest voter of some current winner or just the globally heaviest voter who does not yet vote for $c$. In the latter case we get the greatest additional vote weight for $c$, but in the former case we gain some vote weight for $c$ while simultaneously potentially decreasing the total vote weight that $c$ needs to become a winner. (We say "potentially decrease" since if there are multiple winners then the total vote weight $c$ needs to win won't change. But if we keep on bribing the voters of current winners, this decrease will occur eventually.) Let us consider the following example (see Figure 2). We have candidates $a$, $b$, and $c$ and six voters with weights 1, 2, 2, 2, 3, and 3. Both voters with weight 3 have $a$ as their top candidate, and all the others have $b$ as their top candidate. Thus $a$ receives a total vote weight of 6, $b$ receives a total vote weight of 7, and $c$ receives no votes. If we bribe the two heaviest voters—the two weight-3 voters preferring $a$—then $c$ still loses to $b$. However, if we bribe one weight-3 voter preferring $a$ and one weight-2 voter preferring $b$ then $c$ wins. Examples where bribing the heaviest voters leads to an optimal bribery also exist. This hints that plurality-weighted-bribery may require more than a simple greedy algorithm. Nonetheless, Faliszewski, Hemaspaandra, and Hemaspaandra [FHH06] obtained polynomial-time algorithms for plurality-weighted-bribery and plurality-$bribery.

**Theorem 4.3 ([FHH06])** plurality-weighted-bribery *and* plurality-$bribery *are in* P.

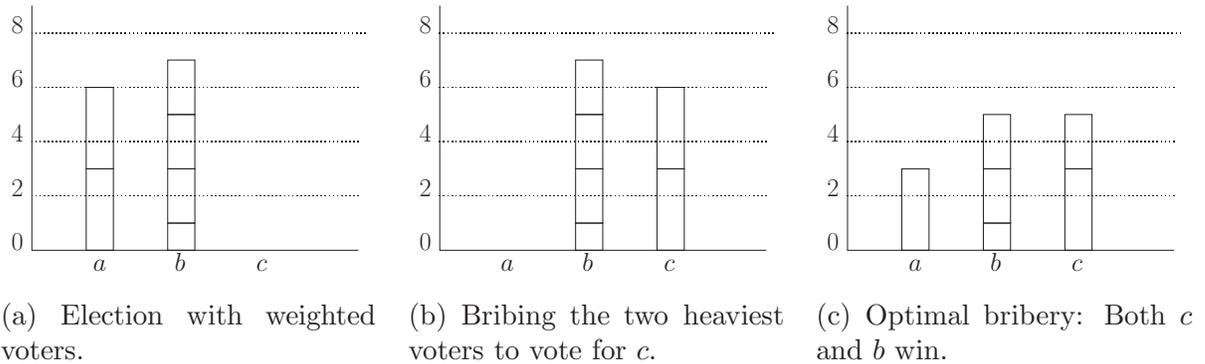

(a) Election with weighted voters.  (b) Bribing the two heaviest voters to vote for $c$.  (c) Optimal bribery: Both $c$ and $b$ win.

Figure 2: Plurality-rule elections where bribing the heaviest voters does not lead to optimal bribery.

Does this mean that bribery for plurality-rule elections is always in P? Again the answer is no. If voters are weighted and have price tags then the problem is NP-complete.

**Theorem 4.4 ([FHH06])** plurality-weighted-$bribery *is* NP-*complete.*

This theorem follows by a fairly simple reduction from the partition problem, which is the problem that asks, given a multiset of $k$ nonnegative integers, whether the multiset can be partitioned into two multisets that each sum to the same value.

Theorems 4.1, 4.2, 4.3, and 4.4 capture the complexity of manipulation and bribery for plurality-rule elections, showing in particular how the complexity of bribery problems eventually jumps to NP-completeness as we make the setting more and more challenging. However, one can pinpoint the jump's location even more precisely. We have been assuming as our default that all the numbers within our problems (i.e., the weights and the prices) are represented in binary. What if we represent these numbers in unary? Let plurality-weighted$_{\text{unary}}$-$bribery be the plurality-weighted-$bribery problem with weights represented in unary and let plurality-weighted-$bribery$_{\text{unary}}$ be the plurality-weighted-$bribery problem with prices encoded in unary. Using a dynamic-programming approach, [FHH06] showed that these problems are in P. What this shows is that the plurality-weighted-$bribery's NP-completeness hangs by the slenderest of threads: Informally put, if either the weights or prices are represented by fairly small numbers, the problem slips into P.

**Theorem 4.5 ([FHH06])** plurality-weighted-$bribery$_{\text{unary}}$ *and* plurality-weighted$_{\text{unary}}$-$bribery *are in* P.

The above results on manipulation and bribery for plurality-rule elections capture the complexity of these problems in many important settings. However, as mentioned in the introduction, a far more satisfying goal is to find some simple rule that determines for which election systems bribery and manipulation problems are easy and for which election systems they are hard. Such general results, which we are going to present now as Theorems 4.7 and 4.10 and Corollary 4.9, are known as dichotomy results.

Conitzer and Sandholm [CS02] observed that, for an election system $\mathcal{E}$ for which the winner problem is in P, if the voters are unweighted and there are a fixed number of candidates then $\mathcal{E}$-manipulation is in P. This result holds because a manipulator can easily evaluate all possible manipulations. The result yields the following corollary (which, though not explicitly stated in [CS02], should naturally be attributed to that paper). For a scoring vector $\alpha$, let $\alpha$-manipulation (respectively, $\alpha$-weighted-manipulation) denote the (weighted) manipulation problem and let $\alpha$-bribery (respectively, $\alpha$-weighted-bribery and $\alpha$-weighted-$bribery) denote the (weighted and weighted-plus-priced) bribery problem with respect to the scoring protocol that uses $\alpha$.

**Corollary 4.6 ([CS02])** *For each scoring vector $\alpha$, $\alpha$-manipulation is in* P.

Can we obtain a sharp, easy-to-use classification result with respect to manipulation for scoring protocols with weighted voters? Conitzer, Lang, and Sandholm [CS02, CLS03] took some first steps in this direction. In particular, they observed that for each $m \geq 3$, $(\overbrace{1,\ldots,1}^{m-1},0)$-weighted-manipulation (this is $m$-candidate veto) and $(m-1, m-2, \ldots, 1, 0)$-weighted-manipulation (this is $m$-candidate Borda count) are NP-complete. (Note that for two candidates both the Borda count and veto are equivalent to plurality-rule elections, since they all have the same scoring vector, $(1,0)$.) Although certainly interesting, these results don't reach the goal of classifying the complexity of weighted manipulation for all scoring protocols. The problem of full classification was recently solved by Hemaspaandra and Hemaspaandra [HH05], who obtained the following dichotomy theorem for scoring protocols with respect to $\alpha$-weighted-manipulation. (The 3-candidate case—and some other cases—of the Hemaspaandra–Hemaspaandra manipulation dichotomy work has been independently obtained by Procaccia and Rosenschein [PR06]. The 3-candidate special case has also been independently obtained in an unpublished manuscript of Conitzer, Sandholm, and Lang [CSL05].)

**Theorem 4.7 ([HH05])** *Let $\alpha = (\alpha_1, \ldots, \alpha_m)$ be a scoring vector. If $\alpha_2 = \cdots = \alpha_m$ then $\alpha$-weighted-manipulation is in* P. *In all other cases, this problem is* NP-*complete.*

This result clarifies a few things. In particular, it shows that plurality-rule elections are in fact quite special among scoring protocols, and it shows why scoring protocols tend to jump to NP-completeness at 3 candidates (in particular, the results on veto and Borda count mentioned in the previous paragraph are special cases of Theorem 4.7). Due to space limitations, we omit the proof of Theorem 4.7, which proceeds by a reduction from the partition problem.

Theorem 4.7 is a crisp, natural example of how one can obtain complete characterization results (admittedly, with respect to scoring protocols) regarding the computational complexity of manipulation. It would be great to be able to translate this result from the context of manipulation to that of bribery, in the hope of getting a complete characterization result for bribery. A natural first step would be an attempt to prove, for example, that all bribery problems for each given election system are at least as hard as the respective manipulation problems. Unfortunately, if we want to capture all possible election systems then such a result is impossible. For example, for approval voting the manipulation problem is in P (all manipulators simply approve of just the candidate they are seeking to make win, and no one else), but bribery for approval voting is NP-complete [FHH06]. On the other hand, [FHH06] also constructs an (artificial) election system in which the opposite happens: The bribery problem is in P and the manipulation problem is NP-complete. However, if we stay in the realm of scoring protocols then some extremely useful translations from manipulation to bribery are possible.

First, one can observe that if voters have price tags then bribery is just a generalized manipulation. (As an easy exercise, the reader is encouraged to show that this holds.) The following theorem is a slightly weakened version of a result from [FHH06].

**Theorem 4.8 ([FHH06])** *For each scoring vector $\alpha$, $\alpha$-weighted-manipulation is $\leq_m^p$-reducible to $\alpha$-weighted-$bribery.*

Taking Theorems 4.7 and 4.8 together and by inspecting the reduction that underlies the proof of Theorem 4.4, we can obtain the following corollary.

**Corollary 4.9 ([FHH06])** *Let $\alpha = (\alpha_1, \ldots, \alpha_m)$ be a scoring vector. If $\alpha_1 = \cdots = \alpha_m$ then $\alpha$-weighted-$bribery is in* P. *In all other cases, this problem is* NP-*complete.*

Theorem 4.8 is not entirely satisfactory. Although it translates results on manipulation problems to results on bribery problems, this translation comes at the cost of introducing price tags for voters. In fact, a much stronger translation can be obtained. In particular, using the proof of Theorem 4.7, with much work and problem-reduction trickery, the following dichotomy theorem for weighted bribery with respect to scoring protocols can be shown.

**Theorem 4.10 ([FHH06])** *Let $\alpha = (\alpha_1, \ldots, \alpha_m)$ be a scoring vector. If $\alpha_2 = \cdots = \alpha_m$ then $\alpha$-weighted-bribery is in* P. *In all other cases, this problem is* NP-*complete.*

Note that this theorem essentially replaces the word "manipulation" in Theorem 4.7 with the word "bribery." However, achieving this replacement is far from trivial. The proof follows by first observing that the reduction used in [HH05] can be tweaked to, instead of mapping from the partition problem, map from a restricted version of the partition problem that in effect causes the reduction to produce instances of manipulation problems with certain very special properties. These properties ensure that instances of these manipulation problems can, almost verbatim, be interpreted as instances of the analogous bribery problems.

The above discussion presents results that interrelate bribery and manipulation, and we have seen that doing so helps us obtain broad dichotomy results for bribery. We conclude by mentioning some open problems regarding bribery and manipulation.

One open direction is to seek dichotomy results whose range of applicability is broader than the class of scoring protocols. As mentioned earlier, [HH05, Section 3] already handles uniform *families* of scoring protocols. However, one may hope for even more broadly applicable results. A second open direction is to consider approximation algorithms. This makes sense, for example in the case of bribing priced voters. We would certainly like to know what the cheapest way is of making our preferred candidate a winner by bribery, but we would also be quite satisfied with a cost that—though not the best—is close to optimal. Are there approximation algorithms for plurality-weighted-$bribery or $\alpha$-weighted-$bribery, where $\alpha$ is some scoring vector? If there are such approximation algorithms for these problems then perhaps there even are polynomial-time approximation schemes. Ideally, we would like to obtain a dichotomy result that crisply classifies each scoring protocol as having or not having a polynomial-time approximation scheme. We hope this section will serve as an invitation to the reader to tackle these open problems.

# 5 Complexity of Control: Making Someone Win or Keeping Someone From Winning

The previous section covered manipulation and bribery. Although manipulation and bribery are somewhat different issues, they both have the property that only the voted preferences are changed. The structural properties of the election are not changed.

In real life, however, many attempts to influence elections work by seeking to change the structural properties of elections. By structural changes, we refer to such actions as adding candidates, deleting candidates, adding voters, deleting voters, partitioning candidates, and partitioning voters. The term *control* is used to describe issues related to influencing an election's outcome by changing its structure.

We mention in passing that many real-world attempts to influence elections are attempts to simultaneously influence the structure of an election and influence the way voters vote. For example, when an advertisement for candidate $c$ appears on television, it may be simultaneously trying to get voters who most favor $d$ to switch to $c$, and to get people who already most prefer $c$ but weren't planning on voting to make the effort to go and vote. However, research papers on complexity typically study manipulation and control issues separately. The study of bribery is somewhat of an exception, as bribery, though akin to manipulation, is an atypically flexible form of manipulation, due to the manipulated voters not being fixed as part of the problem input. For this reason, to many people (including the authors) bribery feels somewhat control-like in addition to being very manipulation-like.

For reasons of space, this section will cover control, but with very little stress on formality or even on stating results individually, and instead will simply present an informal discussion about control. We will particularly try to point out what the real-life inspirations are for each type of control. We should warn the reader that in doing so we are taking liberties. For example, we will use as examples some recent American presidential elections. However, in reality, American presidential elections operate under a subtle and obscure system (deeply related, in fact, to partitioning of voters) known as the Electoral College, rather than by direct election by plurality rule. In our informal examples, we will often willfully ignore this and speak as if a presidential election were simply a big plurality-rule election.

We will often discuss both the constructive case—seeking to make a preferred candidate (uniquely) win—and the destructive case—keeping a despised candidate from being a (unique) winner. The complexity of constructive control was first studied in a seminal paper of Bartholdi, Tovey, and Trick [BTT92]. The study of destructive control was initiated much more recently, namely, in work of Hemaspaandra, Hemaspaandra, and Rothe [HHR05]. We mentioned in the introduction that in some subareas of electoral research the focus is on winning and in some the focus is on being the unique winner. For the study of control—in both the constructive and destructive cases—the focus has always been on the case of making a candidate be, or not be, a *unique* winner. Thus, throughout this section, when speaking of our problems or referring to results, when we say (and for brevity and grace we will always just say) winner/wins/winning/etc., we *always* implicitly mean unique winner/uniquely wins/uniquely winning/etc. (the only exception regards the paragraph below on tie-handling rules for subelections, since that directly addresses what happens in subelections when there are tied winners). It is very important to keep this shorthand in mind, since in this section when we say things such as "you can tell whether a despised candidate can be precluded from winning," we always mean "you can tell whether a despised candidate can be precluded from being the unique winner (namely, by either not being a winner at all or by being part of a group of two or more winners)."

Let us start with the issue of *control by adding candidates*. Formally viewed as a set (as all these problems are when seeking rigorous results), this becomes, with respect to some election system $\mathcal{E}$, the sets $\mathcal{E}$-constructive-control-by-adding-candidates and $\mathcal{E}$-destructive-control-by-adding-candidates. The former is defined as follows.

**Name:** $\mathcal{E}$-constructive-control-by-adding-candidates.
**Given:** A set $C$ of original candidates, a pool $D$ of potential additional candidates, a distinguished candidate $c \in C$, and a set $V$ of voters with preferences over $C \cup D$.
**Question:** Is there a set $D' \subseteq D$ such that, under election system $\mathcal{E}$, $c \in C$ is a winner of the election having candidates $C \cup D'$ with the voters being $V$ with the preferences of $V$ restricted to $C \cup D'$?

That is, can we add some of the additional candidates and by doing so make $c$ a winner? As a real-life motivating example, regarding the 2000 American presidential election, if one wanted George W. Bush to win (and Ralph Nader was not at that time running) one might have chosen to add the candidate Ralph Nader if one believed that that would split voters away from Al Gore and achieve one's desired outcome. $\mathcal{E}$-destructive-control-by-adding-candidates is defined with the identical "Given" field, but its question regards, naturally, not trying to make a preferred candidate win but rather making a despised candidate not win:

**Name:** $\mathcal{E}$-destructive-control-by-adding-candidates.
**Given:** A set $C$ of original candidates, a pool $D$ of potential additional candidates, a distinguished candidate $c \in C$, and a set $V$ of voters with preferences over $C \cup D$.
**Question:** Is there a set $D' \subseteq D$ such that, under election system $\mathcal{E}$, $c \in C$ is not a winner of the election having candidates $C \cup D'$ with the voters being $V$ with the preferences of $V$ restricted to $C \cup D'$?

The same real-life motivating example works here, except shifted to the case of focusing on an organization who despised Gore and wanted simply to see him not win.

From here on, we often won't formally describe the problems as sets, but will leave the descriptions very informal (even though in our results table we will refer to the formal sets). Interested readers can find the detailed, formal descriptions in [BTT92,HHR05]. We will also, until the results table, stop mentioning $\mathcal{E}$ explicitly.

Just as one can study control by adding candidates, one can similarly study *control by deleting candidates*. The input is $C$, $V$, $c \in C$, and a natural number $k$, and in the constructive case one wants to know whether by deleting at most $k$ candidates one can make $c$ a winner, and in the destructive case one wants to know whether by deleting at most $k$ candidates (with the deletion of $c$ forbidden) one can ensure that $c$ is not a winner. The same motivating example as above works here. For example, for the constructive case, in both the 2000 and 2004 American presidential elections, some people who wanted Al Gore or John Kerry to win sought to convince/urge/pressure Ralph Nader to withdraw from the race. (Regarding the destructive case, many people whose view was "Anyone but Bush" also naturally wanted Ralph Nader to withdraw.)

Turning to the problems of control by adding voters and control by deleting voters (typically treated as separate, though in the real world these issues interact), in *control by adding voters* (respectively, *control by deleting voters*), one asks whether by adding at most $k$ from a pool of additional voters (respectively, by removing at most $k$ of the initial voters) a given election will make $c$ a winner (constructive case) or not a winner (destructive case). A real-world motivation here for considering the case of adding voters is so-called get-out-the-vote efforts. A political party, on the day of an election, might send vans to bring to the polls voters who the party believes favor its candidate but who might without the vans not make the effort to show up and vote. Or a political party might air ads designed to energize some part of its base and get them to decide to show up and vote (e.g., by putting an ad in CACM saying "That other party's candidate, Marty Meanie, if elected will put a tax on every line of code, with a surtax on comment lines. Only by voting can you help prevent that terrible future!"). One real-world motivation for having the case of deleting voters is less openly admitted by parties and groups today, but is widely viewed as occurring: vote suppression efforts. A party might run ads designed to sap the will to get-out-to-vote of the base of its key opponent. Or if one were a media outlet favoring Gore in 2000 and one went on the air and called Florida for Gore while voting was still going on in the more conservative Panhandle part of Florida, that might lead voters in that part of the state not to show up and cast their (more conservative) votes, since they would believe that the state was already a lost cause (here, we are taking into account the Electoral College structure of that election).

Control by voter addition/deletion is also, in a wider view of affairs, related to disenfranchisement—that is, it is related to the issue of which broad groups are, under law, allowed/not-allowed to vote and what hurdles (sometimes via requirements and sometimes via intimidation) are used to in effect prevent broad groups from voting. Using American history for examples, some cases—ranging in modern-day acceptance from the overwhelmingly accepted to the overwhelmingly deplored—of direct exclusions under law include the facts that (today and in the past) children are not allowed to vote, that (today and in the past) resident aliens are not allowed to vote in most elections, that (today and in the past) felons lose their federal vote for life, that (until 1961) citizens living in the US seat of government weren't represented in the Electoral College (and so didn't influence presidential elections), and that (in the past) women and slaves—and in many Southern states all African-Americans—were not allowed to vote. Some American historical cases of exclusion-in-effect via requirements include poll taxes and literacy tests.

For reasons of space, we won't go into detail in defining the various partition schemes, but we mention that partition attempts regarding both voters and candidates occur often in real life. In these schemes, we have more than one voting round, having to do with partitions of the candidates or the voters. We give only examples of the latter, as those are particularly natural. As a first such example, every time an American state legislature does a Congressional redistricting, it may be a type of prepackaged attempt to partition by voters: In a typical redistricting, the dominant party tries to make sure that in as many districts as possible it has enough supporters to hold the seat but not so many supporters as to waste their votes by winning that seat with too much support. (The side that doesn't control the legislature usually refers to such redistricting via the pejorative term "gerrymandering.") As another example of partition by voters, in some American states in elections for various state-wide officials—say, for their US senator—candidates are chosen by separate party primaries in which only members of each given party vote. Then only the winners of those primaries participate in the final election, in which all the registered voters can vote (to make this example work, let us assume that in this state there are no independent voters).

When dealing with partition schemes, one must have some rule as to what happens when there is a tie *in a subelection*. In the results table later, following [HHR05] which first studied these tie-handling models in this context, we use TP ("ties promote") to indicate the rule that all people who tie as winners move forward from subelections, and we use TE ("ties eliminate") to indicate the rule that only unique winners of subelections move forward. Note that the tie-handling rules affect just the subelections, not the final election round of a given partition system (which as is conventional in the study of electoral control always focuses on unique winnership).

This concludes our presentation of the standard types of electoral control. With each existing for both the constructive case and the destructive case, the standard types of control are adding candidates, deleting candidates, adding voters, deleting voters, and, though we did not discuss them in any detail here, three types of partition schemes with each of those three occurring in both the TP and the TE models. So, in brief, there are ten standard types of constructive control, and each of those ten also has a destructive control analog. Each of these twenty control problems is (for each fixed election rule) simply a set. And that set is either computationally easy (meaning it is easy given an instance to decide whether the desired outcome can be achieved using that type of control) or that set is computationally hard (meaning it is hard—say, NP-hard—given an instance to decide whether the desired outcome can be achieved using that type of control).

Indeed, the study of the complexity of electoral control looks at these issues in almost exactly those terms, but with one twist. That twist regards the easy problems. In particular, there are two very different ways a problem might be easy. Consider an election system and a particular type of control for which the type of control at issue can *never* change someone from not being a winner to being a winner within that election system. In that case, the formal control problem (assuming the winner problem—recall that by that we in this section implicitly mean the unique winner problem—for that election system is in P) of course is in P, but for a very uninteresting reason. In that case, we say the problem is *immune* to constructive control: The given type of control can never shift one's preferred candidate from not winning to winning. Immunity to destructive control is defined analogously: The given type of control can never shift one's despised candidate from winning to not winning. If a problem is not immune and is in P, then we say it is *vulnerable*. So when vulnerability holds for a type of control, then that type of control (since immunity does not hold) sometimes actually makes a profound difference, and in polynomial time we can tell whether a given instance is one where the desired constructive or destructive electoral outcome can be achieved. Although knowing that there exists *some way* to achieve the desired outcome is different from knowing some such way, it turns out that for every vulnerability result stated in our forthcoming results table, Table 1, there is an algorithm that not just determines when control can be exerted, but that also gives the exact control actions to take in order to exert the desired control. Finally, if a control problem is not immune and is NP-hard, then we say it is *resistant* to control. Although immunity is the most desirable case (at least if one is not seeking to exert control, but rather is an election-system designer seeking to frustrate those wishing to influence outcomes via control), resistance is also a very desirable case—it means that the general problem of determining whether a given election instance can be controlled is computationally intractable (NP-hard). Although not all NP-hard problems are NP-complete, for every resistance result in our forthcoming results table an NP upper bound is obvious, so each resistance result of the table in fact represents an NP-completeness claim. (We mention in passing that in the literature election systems that are not immune to a given type of control—that are either vulnerable or resistant to it—are said to be *susceptible* to that type of control.)

|  | plurality | | Condorcet | | approval | |
| --- | --- | --- | --- | --- | --- | --- |
| control by | construct. | destruct. | construct. | destruct. | construct. | destruct. |
| adding-candidates | *R* | **R** | *I* | **V** | **I** | **V** |
| deleting-candidates | *R* | **R** | *V* | **I** | **V** | **I** |
| partition-of-candidates | *TE: R* <br> *TP: R* | **TE: R** <br> **TP: R** | *V* | **I** | **TE: V** <br> **TP: I** | **TE: I** <br> **TP: I** |
| run-off-partition-of-candidates | *TE: R* <br> *TP: R* | **TE: R** <br> **TP: R** | *V* | **I** | **TE: V** <br> **TP: I** | **TE: I** <br> **TP: I** |
| adding-voters | *V* | **V** | *R* | **V** | **R** | **V** |
| deleting-voters | *V* | **V** | *R* | **V** | **R** | **V** |
| partition-of-voters | **TE: V** <br> **TP: R** | **TE: V** <br> **TP: R** | *R* | **V** | **TE: R** <br> **TP: R** | **TE: V** <br> **TP: V** |

Table 1: Results on constructive and destructive control. The problem's name is implicitly described by the table, e.g., the top right "V" refers to the case approval-destructive-control-by-adding-candidates. Results due to Bartholdi, Tovey, and Trick [BTT92] are italicized. Results due to Hemaspaandra, Hemaspaandra, and Rothe [HHR05] are in boldface. Key: I = immune, R = resistant, V = vulnerable, TE = ties-eliminate, TP = ties-promote.

We said earlier that we would not stress results, but it certainly makes sense to see what is known. Table 1, which is taken from [HHR05], summarizes results on constructive and destructive control. These results are due to Bartholdi, Tovey, and Trick [BTT92] and Hemaspaandra, Hemaspaandra, and Rothe [HHR05], and are about just the collection of election systems that those papers studied. Regarding open directions, we urge the reader to study the control problems of other election systems and to seek to find in a broader way what it is that makes some control problems computationally easy and some computationally hard.

From Table 1, some interesting observations are clear. There are settings immune to constructive control that are vulnerable to destructive control. Condorcet elections with respect to control by adding candidates is one such example. Perhaps somewhat more surprisingly, there also are settings immune to destructive control that are vulnerable to constructive control. Approval elections with respect to control by deleting candidates are one such example. Quite interestingly, there are settings vulnerable to destructive control yet resistant to constructive control. In these, you may not be able to efficiently tell whether your favorite candidate can be made to win, but you can efficiently tell whether a despised candidate can be precluded from winning. Condorcet elections with respect to control by adding voters is one such example. Also very interesting is that tie-handling rules can make a tremendous difference. For example, for plurality-rule elections with respect to control by partition of voters, vulnerability holds in the "ties eliminate" model but resistance holds in the "ties promote" model.

Finally, the most glaring observation is that for not one of the systems is it the case that resistance-or-immunity to control holds under all the twenty studied control attacks (ten constructive and ten destructive). Each system studied has good properties (immunity or resistance) under some attacks, but has bad properties (is vulnerable) under other attacks. In fact, at the time the table's work on control was completed (early 2005), no system was proven to be immune-or-resistant to all twenty types of control (or even to the ten constructive types, or to the ten destructive types). However, recently, work of Hemaspaandra, Hemaspaandra, and Rothe [HHR06] has shown how to "hybridize" collections of elections in a way such that the hybrid election has a polynomial-time winner problem if all its constituent systems have polynomial-time winner problems, yet the hybrid system *is resistant to every one of the twenty types of control to which one or more of its constituent systems is resistant*. Simply put, the hybridization scheme combines strengths without adding weaknesses. From that work it now is known that there is an election system (admittedly, an artificial one, since it is built by hybridizing enough systems—some of which had to be constructed just for that purpose—to have, between them, all the right underlying resistances) that is resistant to all twenty standard types of electoral control yet has a polynomial-time winner problem.

## 6 Conclusions

This chapter has surveyed some recent progress in the complexity of elections, focusing primarily on providing an overview of some of the results obtained to date in an ongoing collaborative research project between Düsseldorf and Rochester. The authors firmly believe that the study of elections is a showcase area where interests come together spanning such CS specialties as theory, systems, and AI and such other fields as economics, business, operations research, and political science. And within the study of elections, the central importance of complexity/algorithmic issues has emerged more clearly with each passing year. Complexity offers a nonclassical yet powerful tool to frustrate those who seek to manipulate or control electoral outcomes. Nonetheless, much remains to be learned. In this chapter's sections, we have tried to point out in passing some of the questions that seem to us the most interesting and urgent. We commend these questions, and this entire area of study, to all readers and most especially to those younger readers seeking a research area that is fresh, promising, enjoyable, theoretically well-grounded, and well-connected to societal applications.

**Acknowledgments**   We thank S. S. Ravi and Sandeep Shukla for inviting us to contribute to the Festschrift in honor of Dan Rosenkrantz for which this was written, and we thank Dan Rosenkrantz for his tremendous contributions to the field and his sterling example to all.

*and Data Security*, pages 285–297. Springer-Verlag *Lecture Notes in Computer Science #3570*, 2005.

[ER91]   E. Ephrati and J. Rosenschein. The Clarke tax as a consensus mechanism among automated agents. In *Proceedings of the 9th National Conference on Artificial Intelligence*, pages 173–178. AAAI Press, 1991.

[ER93]   E. Ephrati and J. Rosenschein. Multi-agent planning as a dynamic search for social consensus. In *Proceedings of the 13th International Joint Conference on Artificial Intelligence*, pages 423–429. Morgan Kaufmann, 1993.

[FHH06]   P. Faliszewski, E. Hemaspaandra, and L. Hemaspaandra. The complexity of bribery in elections. In *Proceedings of the 21st National Conference on Artificial Intelligence*, pages 641–646. AAAI Press, July 2006.

[Fis77]   P. Fishburn. Condorcet social choice functions. *SIAM Journal on Applied Mathematics*, 33(3):469–489, 1977.

[FKS03]   R. Fagin, R. Kumar, and D. Sivakumar. Efficient similarity search and classification via rank aggregation. In *Proceedings of the 2003 ACM SIGMOD International Conference on Management of Data*, pages 301–312. ACM Press, June 2003.

[Gib73]   A. Gibbard. Manipulation of voting schemes. *Econometrica*, 41(4):587–601, 1973.

[GJ79]   M. Garey and D. Johnson. *Computers and Intractability: A Guide to the Theory of NP-Completeness*. W. H. Freeman and Company, 1979.

[Hem89]   L. Hemachandra. The strong exponential hierarchy collapses. *Journal of Computer and System Sciences*, 39(3):299–322, 1989.

[HH05]   E. Hemaspaandra and L. Hemaspaandra. Dichotomy for voting systems. Technical Report TR-861, Department of Computer Science, University of Rochester, Rochester, NY, April 2005. Journal version to appear in *Journal of Computer and System Sciences*.

[HH06]   C. Homan and L. Hemaspaandra. Guarantees for the success frequency of an algorithm for finding Dodgson-election winners. In *Proceedings of the 31st*31